\documentclass{nature}
\usepackage{amsmath}    % need for subequations
\usepackage{graphicx}   % for figures
\usepackage{color}
\usepackage[10pt]{moresize}
\usepackage{graphics,graphicx}
\usepackage{amsfonts}
\usepackage{amssymb}
\usepackage{amscd}
\usepackage{amsmath}
\usepackage{enumerate}
\usepackage{epsfig}
\usepackage{subfigure}

\usepackage[colorlinks,breaklinks,linkcolor=blue,anchorcolor=blue,citecolor=blue,urlcolor=blue,
dvipdfm
]{hyperref}
\usepackage{bm,mathrsfs}

\begin{document}

\title{Effect of spin relaxations on the spin mixing conductances for a bilayer structure}
\author{D. X. Li$^{1}$, H. Z. Shen$^{1,2}$, H. D. Liu$^{1,2}$, and X. X. Yi$^{1,2 \dag}$}

\maketitle

\begin{affiliations}
\item
Center for Quantum Sciences and School of Physics,
Northeast Normal University, Changchun 130024, China
\item
Center for Advanced Optoelectronic Functional  Materials
Research, and Key Laboratory for UV Light-Emitting Materials and
Technology of Ministry of Education, Northeast Normal  University,
Changchun 130024, China\\
$^\dag$ To whom correspondence should be addressed. E-mail:
yixx@nenu.edu.cn
\end{affiliations}
\baselineskip24pt

\maketitle

\begin{abstract}
The spin current can result in a spin-transfer torque in the normal-metal(NM)|ferromagnetic-insulator(FMI) or normal-metal(NM)|ferromagnetic-metal(FMM) bilayer. In the earlier study on this issue, the spin relaxations were ignored or introduced phenomenologically.
In this paper, considering the FMM or FMI with spin relaxations  described by a non-Hermitian Hamiltonian,  we derive an effective spin-transfer torque and an effective spin mixing conductance in the non-Hermitian bilayer. The dependence of the effective spin mixing conductance on the system parameters (such as insulating gap, \textit{s-d} coupling, and layer thickness) as well as the relations between the real part and the imaginary part of the effective spin mixing conductance are given and discussed. We find that the effective spin mixing conductance can be enhanced in the non-Hermitian system. This provides us with the possibility to enhance the spin mixing conductance.
\end{abstract}
\clearpage

\section*{Introduction}
Spin current is a major issue in the field of spintronics, which is intimately associated with many interesting phenomena such as the giant magnetoresistance effect \cite{prb212410ref1}, current-induced magnetization dynamics \cite{prb212410ref2,prb212410ref3}, and the manipulation and transport of spins in small structures and devices \cite{prb144411ref1,prb144411ref2}. Spin currents can be obtained by utilizing the spin Hall effect (SHE) and detected by the inverse spin Hall effect (ISHE) \cite{prb104412ref1,prb104412ref2,1607ref9,1607ref10,prb144411ref3,prb104412ref3}.
By making use of the SHE in a normal metal (NM), such as Pt or Ta, an electric current causes a spin accumulation, or spin voltage. At the transverse edge of the sample it can be converted into a spin current \cite{prb104412ref1,prb104412ref2,prb104412ref4,prb104412ref6,1607ref5,prb104412ref5}.
When a ferromagnetic insulator (FMI) such as Y$_3$Fe$_5$O$_{12}$ (YIG) \cite{prb104412ref7}, or a thin film ferromagnetic metal (FMM) such as Co \cite{prb104412ref8,prb104412ref9, prb104412ref10,prb104412ref11,prb104412ref12} is combined with the edge of the NM, the SHE spin current flows towards the interfaces, where it can be absorbed as a spin-transfer torque (STT) on the interface \cite{prb212410ref2,prb212410ref3}. The STT influences the magnetization damping or changes the magnetization \cite{prb144411ref4,prb104412ref9, prb104412ref10}. Hence, STT that describes the  interaction
between the spin of the conduction electrons and a localized magnetic moment \cite{1465mainref1} is also a hot topic in spintronics. The spin-transfer torque at the NM/FMI or NM/FMM interface is governed by the spin mixing conductance $G_{r,i}$ \cite{prb144411ref16,prlref21}. And the prediction of  large $G_{r,i}$ for interfaces of YIG with simple metals by first-principles calculations has been
confirmed by experiments \cite{prb144411ref17,prb144411ref18}.

At present, it is significant to find a method to enhance the spin mixing conductance, which would help achieving  magnetic memory devices with more efficient magnetization switching and lower power consumption. A  minimal model for the STT in a NM/FMI and NM/FMM bilayer based on quantum tunneling of spins \cite{prl217203,prb104412} shown  that the spin mixing conductance  is strongly influenced by generic material properties such as interface $s-d$ coupling, insulating gap, and thickness of the ferromagnet, but it slightly depends on the spin relaxation introduced phenomenonly in the spin expectation value.

As we known, quantum systems undergo decoherence due to unavoidable couplings to  environment. As a consequence, the macroscopic quantum superpositions are strongly suppressed and classical behaviour emerges from the quantum regime. The study of open quantum system has been received enormous attention due to its ubiquitous application in
developing quantum information devices, quantum computation and cryptography. However, in  most papers related to spin transfer in bilayer(e.g., NM-FMM bilayer and NM-FMI bilayer),  the FMM or FMI layer is considered much thinner than its spin relaxation length,
such that the spin relaxation can be ignored. In Ref.~\cite{prl217203}, the issue of spin relaxation was studied  phenomenologically by introducing an exponentially decayed factor to the spin expectation value.
They found that the spin mixing conductance  does  not crucially depend on spin relaxation. These give rise to a question that what the role played in a microscopical theory by the spin-relaxation  in the spin transfer? And if the relaxation can play a positive role in the spin transfer?

In recent years, more and more interests have been devoted to study non-Hermitian
Hamiltonians \cite{pra052502ref6,pra063412ref16,pra052502ref8,pra052502ref14,pra052502ref5,pra063412ref14,pra052502ref9,pra052502ref10,pra052502ref8_2,pra052502ref15}. And some attention has been given to situations where a non-Hermitian system interacts with the world of Hermitian quantum mechanics.
For instance, a non-Hermitian analogue
of the Stern-Gerlach experiment has been examined, in which the role of
the intermediate inhomogeneous magnetic field flipping the spin is taken over by an apparatus described by a non-Hermitian Hamiltonian \cite{prd125003ref10}.

This motivates us to  consider a multi-layer with NM described by a Hermitian Hamiltonian and FMI or FMM  by a non-Hermitian Hamiltonian. In the non-Hermitian system, we still  utilize a minimal formalism for the STT based on quantum tunneling of spins. We will derive an effective spin transfer torque in the non-Hermitian system and obtain an effective spin mixing conductance of the non-Hermitian system by the  Landau-Lifshitz (LL) dynamics \cite{prb212410ref3}. Furthermore, we investigate the dependence of the effective spin mixing conductance on the system parameters as well as the relations between the real part and the imaginary
part of the effective spin mixing conductance. The enhancement of the effective spin mixing conductance in the non-Hermitian system is found.

\section*{Results}
\subsection{NM/FMI bilayer described by non-Hermitian system}
\begin{figure}
\begin{centering}
\includegraphics[width=10 cm]{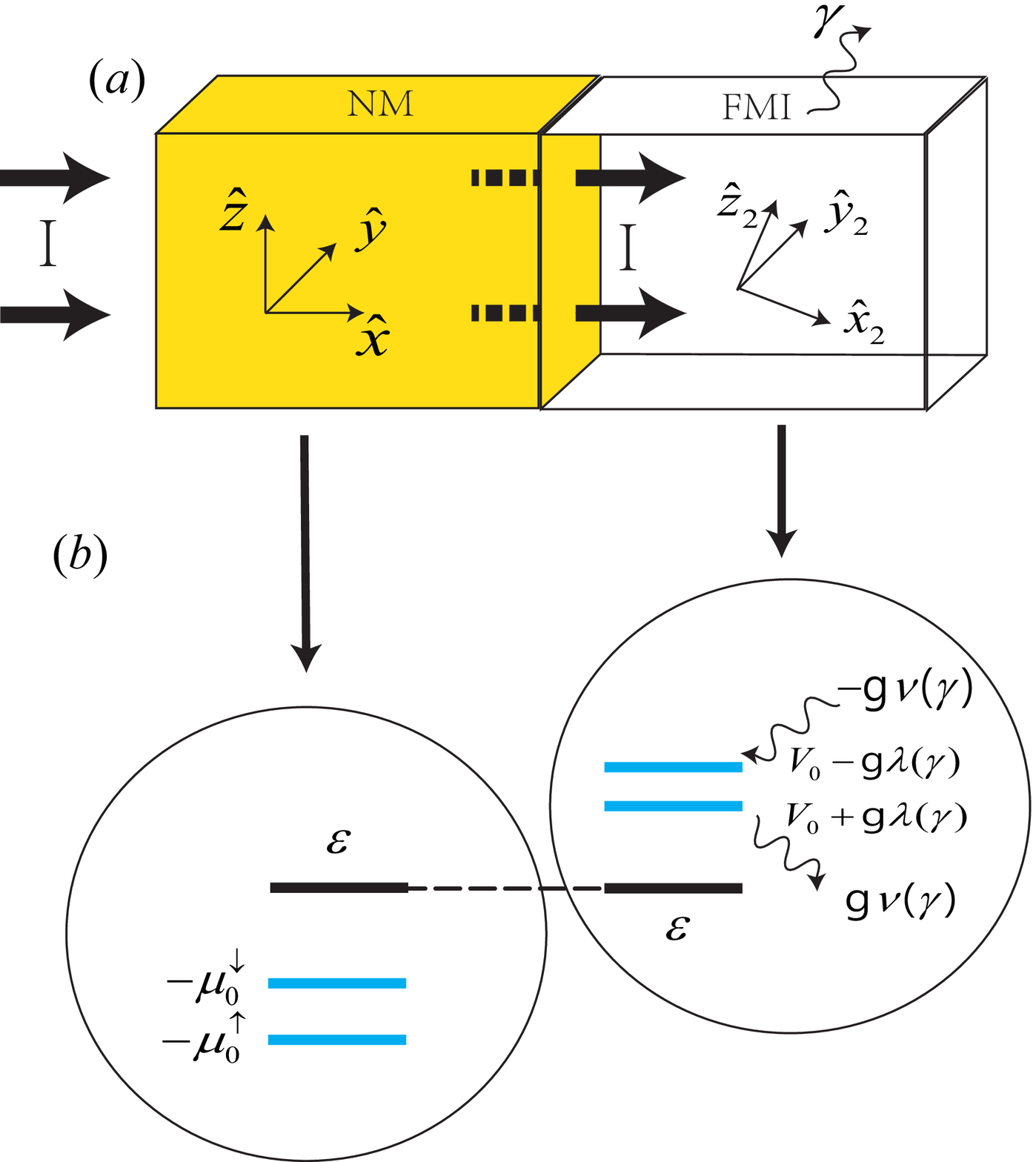}
\par \end{centering}
\caption{\label{FMImodel}(a) Schematic illustration  of the NM/FMI bilayer.   $\gamma$ denotes the spin relaxation rate. (b) The energy levels  near the NM/FMI interface are plotted for  the case of $\nu(\gamma)>0$. Note that all  energy levels for FMI are larger than the NM except the lowest one.}
\end{figure}

A NM/FMI bilayer considered in this paper is shown  in Fig.~\ref{FMImodel}. The normal metal (NM)  at $-\infty<x<0$ is described by $H_L=p^2/2m-\mu_x^\sigma$, where $\mu_x^\sigma=\pm|\bm\mu_x|/2$ is the spin voltage with  $\sigma={\uparrow,\downarrow}$ at position $x$ caused by the SHE \cite{prb104412ref4,prb104412ref6,prb104412ref5}. {The $I$ denotes the current flowing from left to right.} For an up spin incident from the left, the wave function near the interface of left side can be written as,
\begin{equation}\label{leftwavefunction}
\begin{aligned}
|\psi_L(x)\rangle=(Ae^{ik_{0\uparrow}x}+Be^{-ik_{0\uparrow}x})\left( \begin{array}{c} 1\\0 \end{array} \right)+Ce^{-ik_{0\downarrow}x}\left( \begin{array}{c} 0\\1 \end{array} \right),
\end{aligned}
\end{equation}
where $A$, $B$ and $C$ are coefficients to be determined.  $k_{0\sigma}=\sqrt{2m(E+\mu_0^\sigma)}/\hbar$, and $E$ is the Fermi energy. We will consider the electrons moving in $\hat n_x$ direction that have $\bm\sigma\|\hat n_z$ and experience  a positive spin voltage $\bm\mu_0\|\hat n_z$ at the interface.

A ferromagnetic insulator (FMI) at $0<x<\infty$  is described by $H_R=p^2/2m+V_0+g\bm{S}\cdot\bm{\sigma}-i\gamma\sigma_z$, where $V_0>E$ is the potential step. The nonzero $\gamma$ term in the Hamiltonian is introduced to describe the spin relaxation, it may lead to  gain  or loss as we will show later. $\bm{S}=S(\sin\theta\cos\varphi,\sin\theta\sin\varphi,\cos\theta)$ is {the localized magnetization }of FMI and $\bm\sigma$ is the Pauli matrices. In order to describe that the magnetization $\bm{S}$ has a trend towards alignment with the conduction electron spin $\bm\sigma$, we consider $g<0$. The term $g\bm{S}\cdot\bm{\sigma}-i\gamma\sigma_z$ can be rewritten as,
\begin{equation}\label{HRmatrix}
\begin{aligned}
g\bm{S}\cdot\bm{\sigma}-i\gamma\sigma_z=gS'\left( \begin{array}{cc}
\cos\theta' & \sin\theta'e^{-i\varphi}\\
\sin\theta'e^{i\varphi} & -cos\theta'
\end{array} \right),
\end{aligned}
\end{equation}
where
\begin{equation}\label{HRparameter}
\begin{aligned}
&\cos\theta'=\frac{S\cos\theta-i\gamma/g}{S'},\\
&\sin\theta'=\frac{S}{S'}\sin\theta,\\
&S'=\sqrt{S^2\sin^2\theta+(S\cos\theta-i\gamma/g)^2}.
\end{aligned}
\end{equation}
$S'$ in general takes a complex value  which can be written as $S'=\lambda(\gamma)+i\nu(\gamma)$. Note that the energy levels for the NM and FMI are different,  see Fig.~\ref{FMImodel}(b).
The eigenvalues of Eq.~(\ref{HRmatrix}) are $\pm S'$ and the evanescent wave function near the interface of right side that is a superposition of  the right eigenstates of Eq.~(\ref{HRmatrix}) takes the form,
\begin{equation}\label{righteigen}
\begin{aligned}
|\psi_R(x)\rangle=&De^{-q_+x}\left( \begin{array}{c}
\cos{\frac{\theta}{2}'}e^{-i\varphi/2}\\ \sin{\frac{\theta}{2}'}e^{i\varphi/2}
\end{array} \right)+Ee^{-q_-x}\left( \begin{array}{c}
-\sin{\frac{\theta}{2}'}e^{-i\varphi/2}\\ \cos{\frac{\theta}{2}'}e^{i\varphi/2}
\end{array} \right).
\end{aligned}
\end{equation}
Similarly, the evanescent wave function which can be  expanded by  the left eigenstates of Eq.~(\ref{HRmatrix}) is,
\begin{equation}\label{lefteigen}
\begin{aligned}
\langle\hat\psi_R(x)|=&D'e^{-q_+ x}\left( \begin{array}{cc}
\cos{\frac{\theta}{2}'} e^{i\varphi/2}& \sin{\frac{\theta}{2}'} e^{-i\varphi/2}
\end{array} \right)+E'e^{-q_- x}\left( \begin{array}{cc}
-\sin{\frac{\theta}{2}'} e^{i\varphi/2}& \cos{\frac{\theta}{2}'} e^{-i\varphi/2}
\end{array} \right),
\end{aligned}
\end{equation}
where $D, E, D', E'$ are parameters to be determined.
$q_\pm=\sqrt{2m(V_0\pm g S'-E)}/\hbar$ and we restrict ourself to consider the case of $Re(q_\pm)=Re(\sqrt{2m(V_0\pm g S'-E)}/\hbar)>0.$

 {Consider a transparent interface} and recall  the boundary conditions \cite{pra023415ref28, pra052502ref5,PhysRevD.76.125003,PhysRevD.78.025026,pra062116,pra052109} for non-Hermitian system,
\begin{equation}\label{boundary}
\begin{aligned}
|\psi_L(0)\rangle&=|\psi_R(0)\rangle\\
\langle \psi_L(0)|&=\langle \hat\psi_R(0)|,\\
\frac{d}{dx}|\psi_L(x)\rangle\Big|_{x=0}&=\frac{d}{dx}|\psi_R(x)\rangle\Big|_{x=0},\\
\frac{d}{dx}\langle \psi_L(x)|\Big|_{x=0}&=\frac{d}{dx}\langle \hat\psi_R(x)|\Big|_{x=0},
\end{aligned}
\end{equation}
we can obtain the coefficients  in  Eq.~(\ref{righteigen}) and Eq.~(\ref{lefteigen}),
\begin{equation}\label{waveparameter}
\begin{aligned}
&D=\frac{2n_{\downarrow+}\cos\frac{\theta}{2}'Ae^{i\varphi/2}}{\gamma_\theta},
D'=\frac{2n_{\downarrow+}'\cos\frac{\theta}{2}' A^\ast e^{-i\varphi/2}}{\gamma_\theta'},\\
&E=-\frac{2n_{\downarrow-}\sin\frac{\theta}{2}'Ae^{i\varphi/2}}{\gamma_\theta},
E'=-\frac{2n_{\downarrow-}'\sin\frac{\theta}{2}' A^\ast e^{-i\varphi/2}}{\gamma_\theta'},\\
&n_{\sigma\pm}=\frac{k_{0\sigma}}{k_{0\sigma}+iq_{\pm}},
n_{\sigma\pm}'=\frac{k_{0\sigma}}{k_{0\sigma}-iq_{\pm}},\\
&\gamma_\theta=\frac{n_{\downarrow+}}{n_{\uparrow+}}\cos^2{\frac{\theta}{2}'}+\frac{n_{\downarrow-}}{n_{\uparrow-}}\sin^2{\frac{\theta}{2}'},\\
&\gamma_\theta'=\frac{n_{\downarrow+}'}{n_{\uparrow+}'}\cos^2{\frac{\theta}{2}'}+\frac{n_{\downarrow-}'}{n_{\uparrow-}'}\sin^2{\frac{\theta}{2}'}.
\end{aligned}
\end{equation}
Here we consider $|A|^2=N_F|\bm{\mu_0}|/a^3$ that is attributed to the Fermi surface-averaged spin density at the interface, where $N_F$ is the density of states per $a^3$ at the Fermi surface and $a$ is the lattice constant.

The spin of conduction electrons inside the FMI can be obtained by Eq.~(\ref{righteigen}) and Eq.~(\ref{lefteigen}). We will consider them in the frame ($\hat n_{x_2},\hat n_{y_2},\hat n_{z_2}$), where $\hat n_{z_2} \| \bm{S}$, $\hat n_{y_2}=\hat{\bm\mu}_0\times\hat{\bm{S}}/\sin\theta$,  and $\hat n_{x_2}=\hat{\bm S}\times(\hat{\bm S}\times\hat{\bm\mu}_0)/\sin\theta,$  where the hat sign means corresponding unit vector. In this frame, the magnetization $\bm S=(0,0,S)$ and   Eq.~(\ref{righteigen}) and Eq.~(\ref{lefteigen}) can be rewritten as,
\begin{equation}\label{changespinor}
\begin{aligned}
|\psi_R'(x)\rangle=&De^{-q_+x}\left( \begin{array}{c}
\cos{\frac{\alpha}{2}}\\ \sin{\frac{\alpha}{2}}
\end{array} \right)+Ee^{-q_-x}\left( \begin{array}{c}
-\sin{\frac{\alpha}{2}}\\ \cos{\frac{\alpha}{2}}
\end{array} \right),\\
\langle\hat\psi_R'(x)|=&D'e^{-q_+ x}\left( \begin{array}{c}
\cos{\frac{\alpha}{2}}\\ \sin{\frac{\alpha}{2}}
\end{array} \right)^T+E'e^{-q_- x}\left( \begin{array}{c}
-\sin{\frac{\alpha}{2}} \\ \cos{\frac{\alpha}{2}}
\end{array} \right)^T,
\end{aligned}
\end{equation}
where $\alpha=\theta'-\theta$ was defined and $T$ denotes the transposition.

 {{ In closed system, the spin transfer torque is used to describe the change of the macrospin, ${\bm S}$, which is the description of the magnetization from the localized spins. Because of the conservation of angular momentum, the change of the magnetization from the localized spins is equal to the change of the magnetization from conduction electron spins. So one can calculate the change of the magnetization from conduction electron spins to obtain the spin transfer torque. But in open system, the change of the magnetization from conduction electron spins should be equal to the sum of the change of the macrospin ${\bm S}$ and the spin angular momentum transferred to environment, \textit{i.e.} $\bm\tau=d\bm S/dt+\bm\tau_{E}=d\bm M/dt$, where $\bm M=-\gamma_0\sum_m\frac{\hbar}{2}\langle\bm\sigma_m\rangle$ denotes the magnetization {induced by conduction electron spins} and $\bm\tau_{E}$ describes the angular momentum transferred to environment. $\gamma_0$ is gyromagnetic ratio and the subscript $m$ represents electron at the position labeled by $m$. From the above equation, we define the $\bm\tau$ as an effective spin transfer torque of open system, which is the description of the change of the magnetization from conduction electron spins.}} According to the Heisenberg equation in the frame ($\hat n_{x_2},\hat n_{y_2},\hat n_{z_2}$), we have
\begin{equation}\label{sigmaheisenberg}
\begin{aligned}
\frac{d\bm\sigma}{dt}=&-\frac{i}{\hbar}[\bm\sigma,H_R]\\
=&-\frac{i}{\hbar}[\bm\sigma,g\bm S\cdot\bm\sigma]-\frac{1}{\hbar}[\bm\sigma,\gamma\sigma_z]\\
=&\frac{2g}{\hbar}\bm S\times\bm\sigma+i\frac{2\gamma}{\hbar}\sigma_y\hat n_{x_2}-i\frac{2\gamma}{\hbar}\sigma_x\hat n_{y_2},
\end{aligned}
\end{equation}
where we have used the relation $[\sigma_i,\sigma_j]=2i\varepsilon_{ijk}\sigma_k$. Then the STT can be rewritten as,
\begin{equation}\label{tau}
\bm\tau=\gamma_0g\langle\bm{\bar\sigma}\rangle\times\bm S-i\gamma_0\gamma\langle\bar\sigma_{y_2}\rangle\hat n_{x_2}+i\gamma_0\gamma\langle\bar\sigma_{x_2}\rangle\hat n_{y_2},
\end{equation}
{ where $\langle\bm{\bar\sigma}\rangle\equiv\sum_m\langle\bm\sigma_m\rangle=a^2\int_0^\infty dx\langle\bm\sigma\rangle$, $\langle\bm\sigma\rangle=\langle\hat\psi_R'(x)|\bm\sigma|\psi_R'(x)\rangle$ and $\langle\sigma_{i_2}\rangle=\langle\hat\psi_R'(x)|\sigma_i|\psi_R'(x)\rangle$.} And then the  STT can be written as,
\begin{equation}\label{STT}
\begin{aligned}
&\bm\tau=\gamma_0\langle\bm{\bar\sigma}\rangle\times\bm S_{eff},\\
\end{aligned}
\end{equation}
Note that in the frame ($\hat n_{x_2},\hat n_{y_2},\hat n_{z_2}$), $\bm S_{eff}=(0,0,S_{eff})=(0,0,g S-i\gamma),$   we can rewrite $S_{eff}=|S_{eff}|e^{i\phi}$, where $|S_{eff}|=\sqrt{g^2S^2+\gamma^2}$ and $\tan\phi=-\gamma/g S$. { Expanding  Eq.~(\ref{STT}), we obtain the effective spin mixing conductance $G_r$ and $G_i$, respectively, which are obtained from the effective spin transfer torque of open system and are different from those in closed system,}
\begin{equation}\label{expandSTT}
\begin{aligned}
\bm\tau&=\gamma_0\langle\bm{\bar\sigma}\rangle\times\bm S_{eff}=\gamma_0S_{eff}\langle\bar\sigma_{y_2}\rangle \hat n_{x_2}-\gamma_0S_{eff}\langle\bar\sigma_{x_2}\rangle \hat n_{y_2}\\
&=\gamma_0|S_{eff}|a^2N_F[G_r\bm{\hat S}\times(\bm{\hat S}\times\bm{\mu}_0)+G_i\bm{\hat S}\times\bm{\mu}_0],\\
G_{r,i}&=\int_0^\infty\frac{\langle\sigma_{y_2,x_2}\rangle}{N_F|\bm{\mu_0}|\sin\theta}e^{i\phi}dx.
\end{aligned}
\end{equation}
Substituting Eq.~(\ref{waveparameter}) and Eq.~(\ref{changespinor}) into the Eq.~(\ref{expandSTT}), we finally have,
\begin{equation}\label{conductance}
\begin{aligned}
G_r=&i\frac{2(n_{\downarrow+}' n_{\downarrow-}-n_{\downarrow-}' n_{\downarrow+})S}{a^3\gamma_\theta'\gamma_\theta(q_++q_-)S'}e^{i\phi},\\
G_i=&\Bigg[\left(\frac{n_{\downarrow+}n_{\downarrow+}'\cos^2\frac{\theta}{2}'}{q_+}-
\frac{n_{\downarrow-}n_{\downarrow-}'\sin^2\frac{\theta}{2}'}{q_-}\right)
\frac{2\sin\alpha}{a^3\gamma_\theta'\gamma_\theta\sin\theta}-\frac{2(n_{\downarrow+}' n_{\downarrow-}+n_{\downarrow-}' n_{\downarrow+})S\cos\alpha}{a^3\gamma_\theta'\gamma_\theta(q_++q_-)S'}\Bigg]e^{i\phi},
\end{aligned}
\end{equation}
where $|A|^2=N_F|\bm{\mu}_0|/a^3$ has been  used in the derivation.  {Due to the spin relaxation characterized by the term with $\gamma$, the Hamiltonian is non-Hermitian and the values of the $G_r$ and $G_i$ are complex. As was mentioned above, when we consider the influence of environment, the effective transfer torque consists two parts: (i) The change of the macrospin, which can be described by the real parts of the $G_r$ and $G_i$ in the Eq.~(\ref{conductance}). (ii) The gain or loss of the spin angular momentum by environment, which can be denotes by the imaginary parts of the $G_r$ and $G_i$ in the Eq.~(\ref{conductance}).} {The imaginary parts can be also understood as a delay effect in the spin transfer, which are reminiscent of the complex admittance in a delay circuit using capacitor and inductor.}

\begin{figure}
\begin{centering}
\includegraphics[width=13 cm]{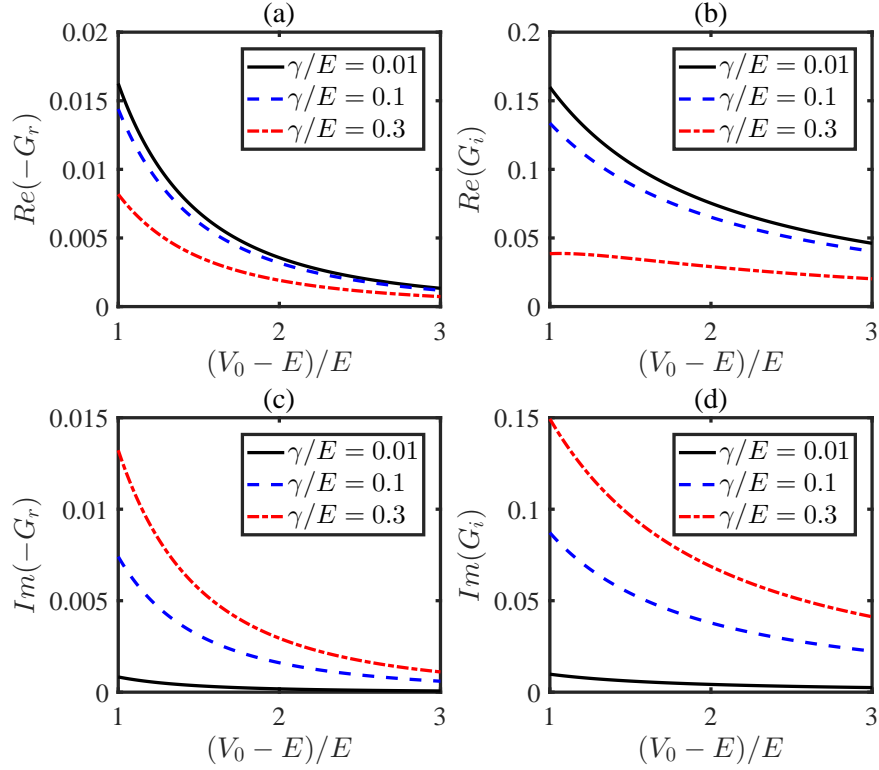}
\par \end{centering}
\caption{\label{FMIgammaSgamma} (a) and (b) show the real parts of $G_r$ and $G_i$ as a function of the insulating gap $(V_0-E)/E$ with different $\gamma/E$, while  (c) and (d) are for the  imaginary parts of $G_r$ and $G_i$. The spin mixing conductance is plotted  in units of $e^2/\hbar a^2$. The other parameters chosen in the figure are $\theta=0.3\pi, -gS/E=0.2, \text{and}~ \mu_0/E=0.01$.}
\end{figure}

In Fig.~\ref{FMIgammaSgamma}, we show the effective spin mixing conductance as a function of insulating gap of the FMI, $(V_0-E)/E$, with different $\gamma$.  The values of the real parts and imaginary parts of $G_r$ and $G_i$ exponentially decay with  the insulating gap of the FMI.  This can be interpreted as that a large insulating gap would result in a short distance of penetrating into the FMI for the electrons.
The spin relaxation can  change neither  the sign of the real parts nor imaginary parts of $G_i$ and $G_r$, implying that the relaxation can not change the direction of the torque.

\begin{figure}
\begin{centering}
\includegraphics[width=16 cm]{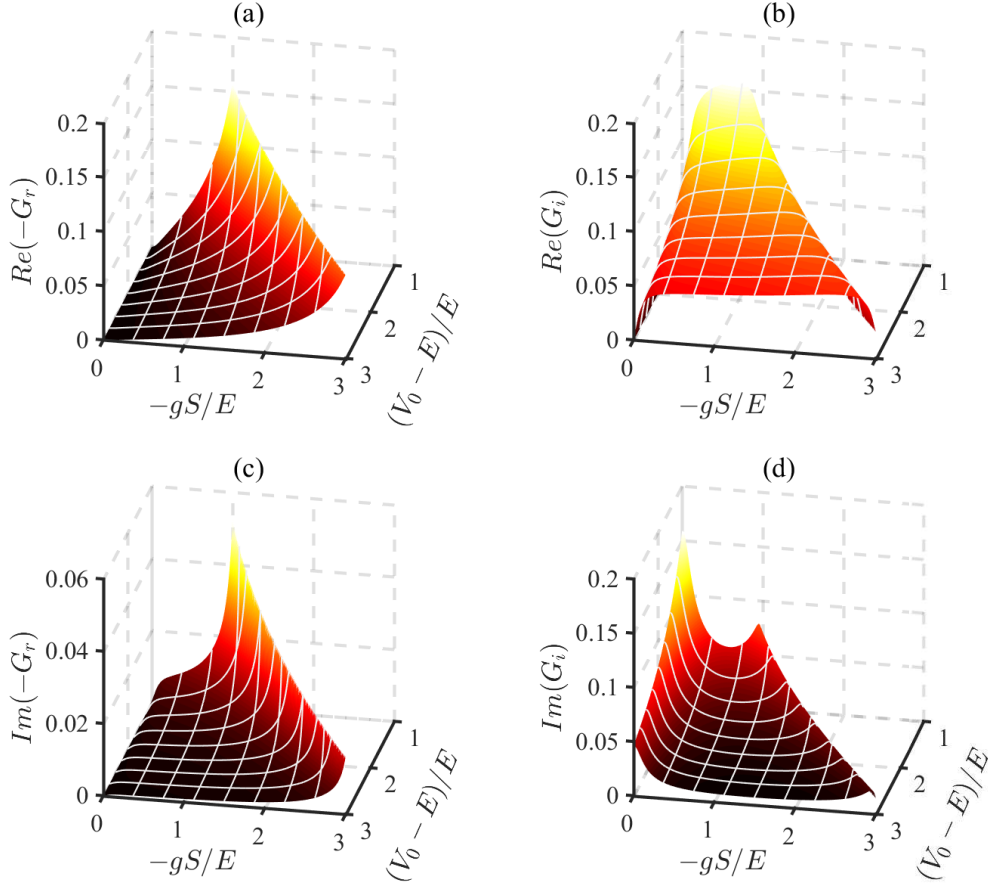}
\par \end{centering}
\caption{\label{FMIGrGi3d}The real parts and the imaginary parts of  $G_{r,i}$ are plotted as a function of the insulating gap $(V_0-E)/E$ and \textit{s-d} coupling $-g S/E$ in FMI. The irrelevant regions where $(V_0-E)/E<-g S/E$ are not plotted. The spin mixing conductance is plotted  in units of $e^2/\hbar a^2$. Here we set $\gamma/E=0.1$ and the other parameters are $\theta=0.3\pi~\text{and}~ \mu_0/E=0.01$. }
\end{figure}

Fig.~\ref{FMIgammaSgamma}  shows that the values of the real parts of  $G_{r,i}$ decrease with $\gamma/E$, while the imaginary parts increases with $\gamma/E$.  To show the dependence of the conductance on both the gap and the \textit{s-d} coupling, we plot the real  and the imaginary parts of $G_{r,i}$ as a function of the insulating gap $(V_0-E)/E$ and  \textit{s-d}  coupling $-g S/E$ in Fig.~\ref{FMIGrGi3d}. The regions where $(V_0-E)/E<-gS/E$ are irrelevant to the problem, so we do not plot these regions in the figure. From Fig.~\ref{FMIGrGi3d}, we can obtain the varying  trends of the conductance with insulating gap $(V_0-E)/E$ and  \textit{s-d} coupling coupling $-g S/E$. In   figure (a) and (c), the varying trends of the real part and imaginary part of $G_r$ are consistent. And in (b) and (d), the variation trends of the real part and imaginary part of $G_i$ are opposite.

\subsection{NM/FMM bilayer described by non-Hermitian system}

\begin{figure}
\begin{centering}
\includegraphics[width=10 cm]{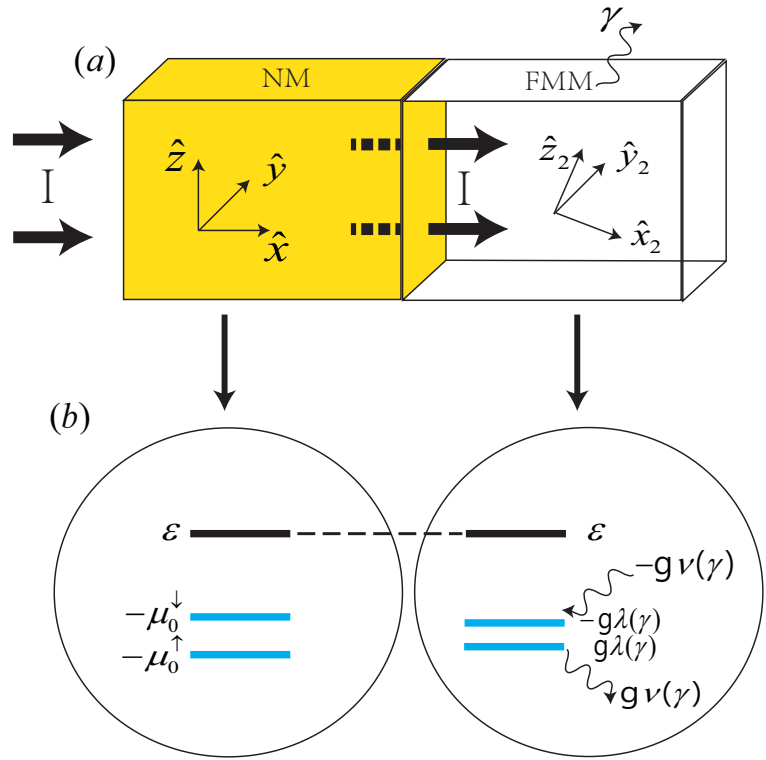}
\par \end{centering}
\caption{\label{FMMmodel}Schematic pictures   similar to   Fig.~\ref{FMImodel}, but for the NM/FMM bilayer. Note that $\nu(\gamma)>0$ in this case. }
\end{figure}

In this section, we will focus on  the effective spin mixing  conductance in the NM/ferromagnetic metal (NM/FMM) bilayer. The models of NM/FMM bilayer similar to   Fig.~\ref{FMImodel} are shown in   Fig.~\ref{FMMmodel}. We consider the bilayer consisting of 2 regions: (1) A NM occupying $-\infty<x<0$ still described by $H_L=p^2/2m-\mu_x^\sigma$ and the wave function is Eq.~(\ref{leftwavefunction}). (2) A FMM in $0<x<l_{FM}$ described by $H_R=p^2/2m+g\bm{S}\cdot\bm{\sigma}-i\gamma\sigma_z$. Following the same procedure as we did in the last section,  we can expand the  wave function with the eigenstates of Eq.~(\ref{HRmatrix}),
\begin{equation}\label{rightlefteigen}
\begin{aligned}
|\psi_R(x)\rangle=&(De^{ik_+x}+Ee^{-ik_+x})\left( \begin{array}{c}
\cos{\frac{\theta}{2}'}e^{-i\varphi/2}\\ \sin{\frac{\theta}{2}'}e^{i\varphi/2}
\end{array} \right)+(Fe^{ik_-x}+Ge^{-ik_-x})\left( \begin{array}{c}
-\sin{\frac{\theta}{2}'}e^{-i\varphi/2}\\ \cos{\frac{\theta}{2}'}e^{i\varphi/2}
\end{array} \right),\\
\langle\hat\psi_R(x)|=&(D'e^{-ik_+ x}+E'e^{ik_+ x})\left( \begin{array}{c}
\cos{\frac{\theta}{2}'} e^{i\varphi/2}\\ \sin{\frac{\theta}{2}'} e^{-i\varphi/2}
\end{array} \right)^T+(F'e^{-ik_- x}+G'e^{ik_- x})\left( \begin{array}{c}
-\sin{\frac{\theta}{2}'} e^{i\varphi/2}\\ \cos{\frac{\theta}{2}'} e^{-i\varphi/2}
\end{array} \right)^T,
\end{aligned}
\end{equation}
where $k_{\pm}=\sqrt{2m(E\mp g S')}/\hbar$ and   $T$ denotes the transposition.  The wave function outside of the bilayer in $l_{FM}<x$ are assumed to vanish for simplicity.

The coefficients can be obtained by matching wave functions and their first derivative at the interface, namely,
\begin{equation}\label{boundary2}
\begin{aligned}
&|\psi_L(0)\rangle=|\psi_R(0)\rangle,|\psi_R(l_{FM})\rangle=0,\\
&\langle\psi_L(0)|=\langle\hat\psi_R(0)|,\langle\hat\psi_R(l_{FM})|=0,\\
&\frac{d}{dx}|\psi_L(x)\rangle\Big|_{x=0}=\frac{d}{dx}|\psi_R(x)\rangle\Big|_{x=0},\\
&\frac{d}{dx}\langle\psi_L(x)|\Big|_{x=0}=\frac{d}{dx}\langle\hat\psi_R(x)|\Big|_{x=0}.\\
\end{aligned}
\end{equation}
According to the boundary conditions Eq.~(\ref{boundary2}), we  obtain,
\begin{equation}\label{parameter2}
\begin{aligned}
&D=\frac{A}{\xi}e^{-ik_+l_{FM}+i\varphi/2}Z_{\downarrow-+}\cos{\frac{\theta}{2}}',E=-e^{2ik_+l_{FM}}D,\\
&D'=\frac{A^\ast}{\xi'}e^{ik_+ l_{FM}-i\varphi/2}Z_{\downarrow-+}'\cos{\frac{\theta}{2}}',E'=-e^{-2ik_+ l_{FM}}D',\\
&F=-\frac{A}{\xi}e^{-ik_-l_{FM}+i\varphi/2}Z_{\downarrow++}\sin\frac{\theta}{2}',G=-e^{2ik_-l_{FM}}F,\\
&F'=-\frac{A^\ast}{\xi'}e^{ik_- l_{FM}-i\varphi/2}Z_{\downarrow++}'\sin\frac{\theta}{2}',G'=-e^{-2ik_- l_{FM}}F',
\end{aligned}
\end{equation}
where
\begin{equation}\label{parameter3}
\begin{aligned}
&Z_{\sigma\alpha\beta}=W_{\sigma\alpha\beta}e^{-ik_\alpha l_{FM}}-W_{\sigma\alpha\bar\beta}e^{ik_\alpha l_{FM}},\\
&Z_{\sigma\alpha\beta}'=W_{\sigma\alpha\beta}e^{ik_\alpha l_{FM}}-W_{\sigma\alpha\bar\beta}e^{-ik_\alpha l_{FM}},\\
&W_{\sigma\alpha\beta}=\frac{k_{0\sigma}+\beta k_\alpha}{2k_{0\sigma}},\\
&\xi=Z_{\uparrow++}Z_{\downarrow-+}\cos^2\frac{\theta}{2}'+Z_{\downarrow++}Z_{\uparrow-+}\sin^2\frac{\theta}{2}',\\
&\xi'=Z_{\uparrow++}'Z_{\downarrow-+}'\cos^2\frac{\theta}{2}'+Z_{\downarrow++}'Z_{\uparrow-+}'\sin^2\frac{\theta}{2}'\\
\end{aligned}
\end{equation}
with $\bar\beta=-\beta$.
According to Eq.~(\ref{expandSTT}) with a replacement $\int_0^\infty\langle\sigma_{x_2,y_2}\rangle dx\rightarrow\int_0^{l_{FM}}\langle\sigma_{x_2,y_2}\rangle dx$, we arrive at
\begin{equation}\label{FMMGri}
\begin{aligned}
&G_r=\frac{\chi(Z_{\downarrow-+}' Z_{\downarrow++}-Z_{\downarrow-+}Z_{\downarrow++}')\sin\theta'}{\xi\xi' a^3\sin\theta}e^{i\phi},\\
&G_i=\frac{e^{i\phi}}{\xi\xi' a^3\sin\theta}\\
&\times\Bigg[\Big(Z_{\downarrow-+}' Z_{\downarrow-+}\mu_+\cos^2\frac{\theta}{2}'-Z_{\downarrow++}' Z_{\downarrow++}\mu_-\sin^2\frac{\theta}{2}'\Big)\sin\alpha+i\chi(Z_{\downarrow-+}Z_{\downarrow++}'+Z_{\downarrow-+}' Z_{\downarrow++})\sin\theta'\cos\alpha \Bigg],
\end{aligned}
\end{equation}
where
\begin{equation}\label{chi}
\begin{aligned}
&\chi=\left( \frac{\sin{\big((k_++k_-)l_{FM}\big)}}{i(k_++k_-)}-\frac{\sin{\big((k_+-k_-)l_{FM}\big)}}{i(k_+-k_-)} \right)\\
&\mu_\pm=2l_{FM}-\frac{\sin{(2k_\pm l_{FM})}}{k_\pm}.
\end{aligned}
\end{equation}
Here  $|A|^2=N_F|\bm{\mu}_0|/a^3$.

\begin{figure}
\begin{centering}
\includegraphics[width=16 cm]{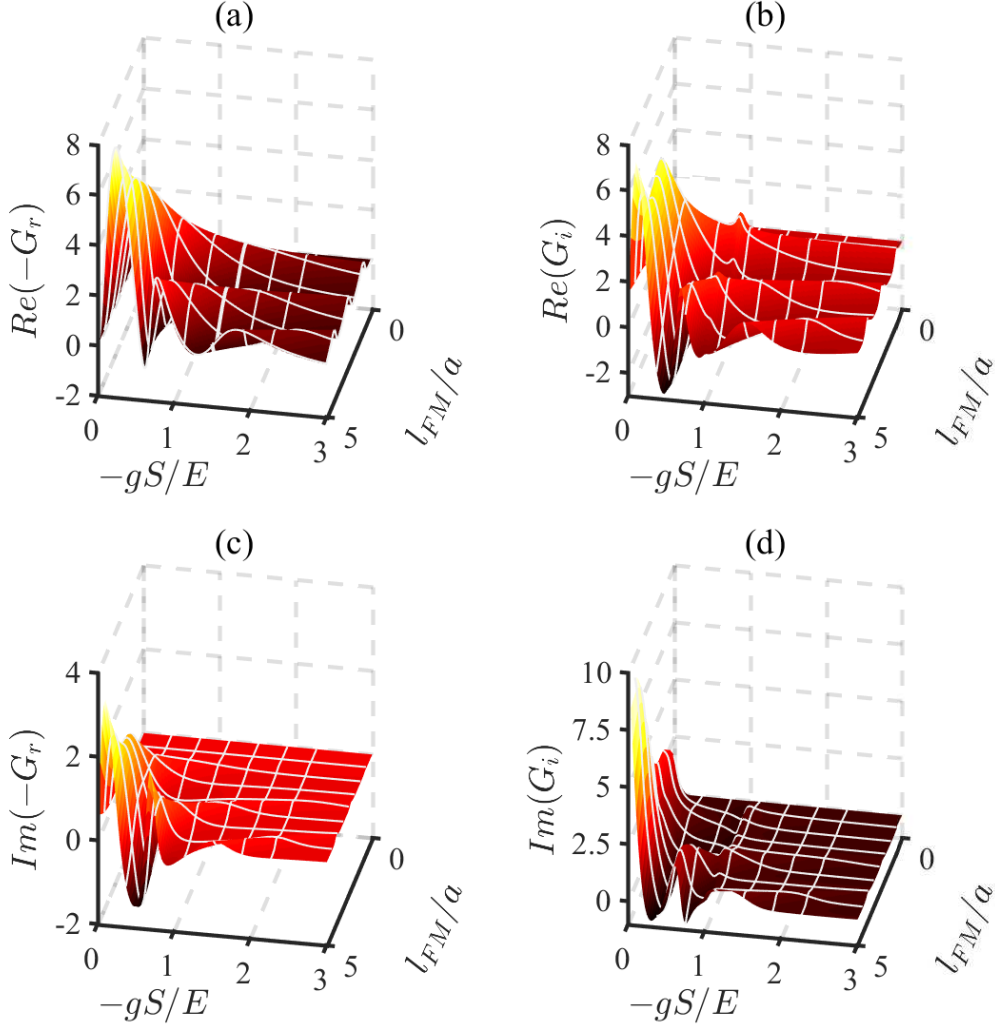}
\par \end{centering}
\caption{\label{FMMGrGi3d}The real parts and the imaginary parts of  $G_{r,i}$ are plotted as a function of the FMM thickness $l_{FM}/a$ and \textit{s-d} coupling $-g S/E$. The units of the spin mixing conductance is $e^2/\hbar a^2$. We set $\theta=0.3\pi,~\gamma/E=0.1,~\text{and}~\mu_0/E=0.01$ for this plot.}
\end{figure}

The real   and imaginary parts of the effective spin mixing conductance as a function of the \textit{s-d} coupling $-g S/E$ and the FMM thickness $l_{FM}/a$ are shown in Fig.~\ref{FMMGrGi3d}. The real parts and the imaginary parts of $G_r$ and $G_i$ decrease nonmonotonically with \textit{s-d} coupling $-g S/E$ and increase nonmonotonically with FMM thickness $l_{FM}/a$. They can change sign by modulating  $-g S/E$ and $l_{FM}/a$. From the figure, we can also find that the conductance (both real and imaginary parts) is a damped-oscillating function.

This might results from  the quantum interference effect when the spin travels into the FMM \cite{prl217203}. The expressions in Eq.~(\ref{FMMGri}) and Eq.~(\ref{chi}) also reveal the oscillatory behavior of $G_{r,i}$ because of the sinusoidal functions of $l_{FM}$ and $k_{\pm}$.

\begin{figure}
\includegraphics[width=16 cm]{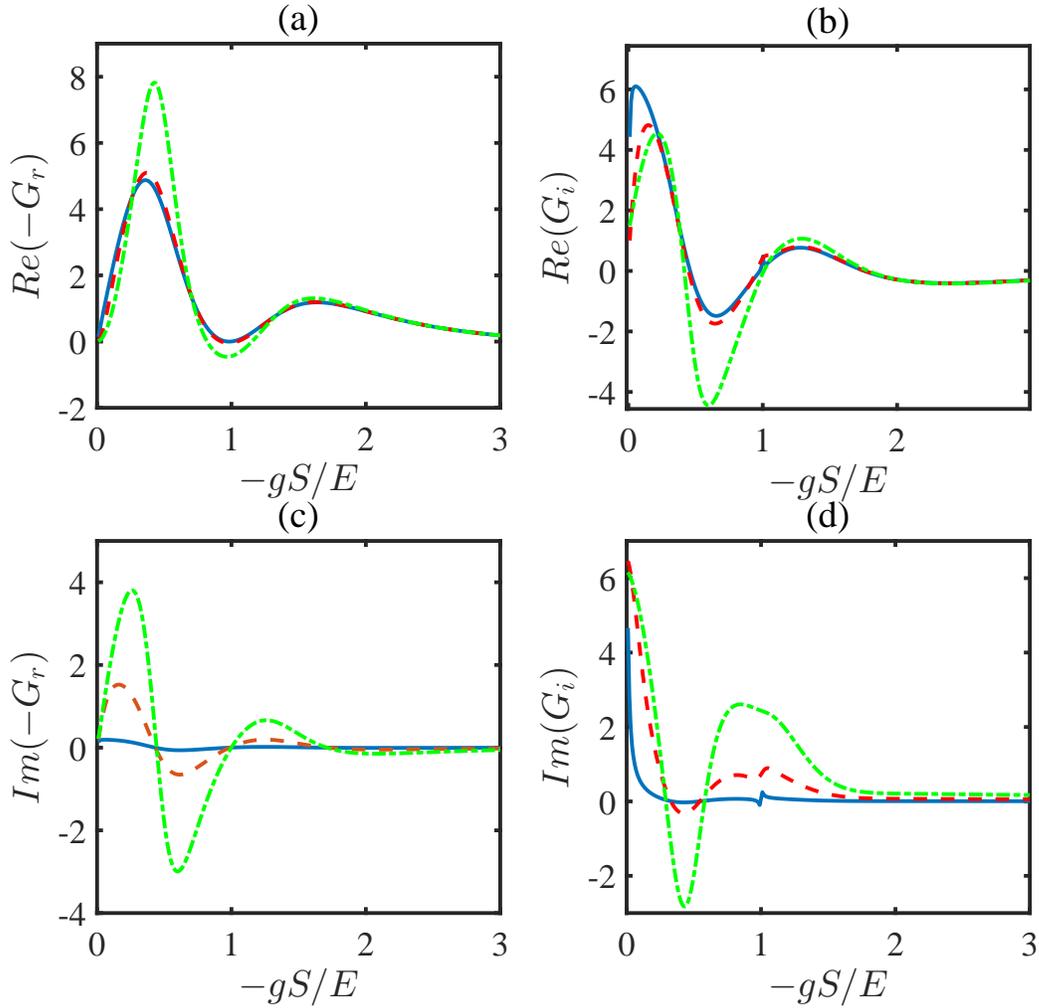}
\caption{\label{FMMGammaS}The real parts and the imaginary parts of $G_r$ and $G_i$ as a function of $-g S/E$ with different values of $\gamma/E$.  $\gamma/E=0.01$ for the   solid line, $\gamma/E=0.1$ for the dashed line, and $\gamma/E=0.3$ for the   dotted line. The other parameters are chosen as $\theta=0.3\pi$, $l_{FM}/a=3.2$, and $\mu_0/E=0.01$. The units for $G_{r,i}$ are also $e^2/\hbar a^2$. }
\end{figure}

\begin{figure}
\includegraphics[width=16 cm]{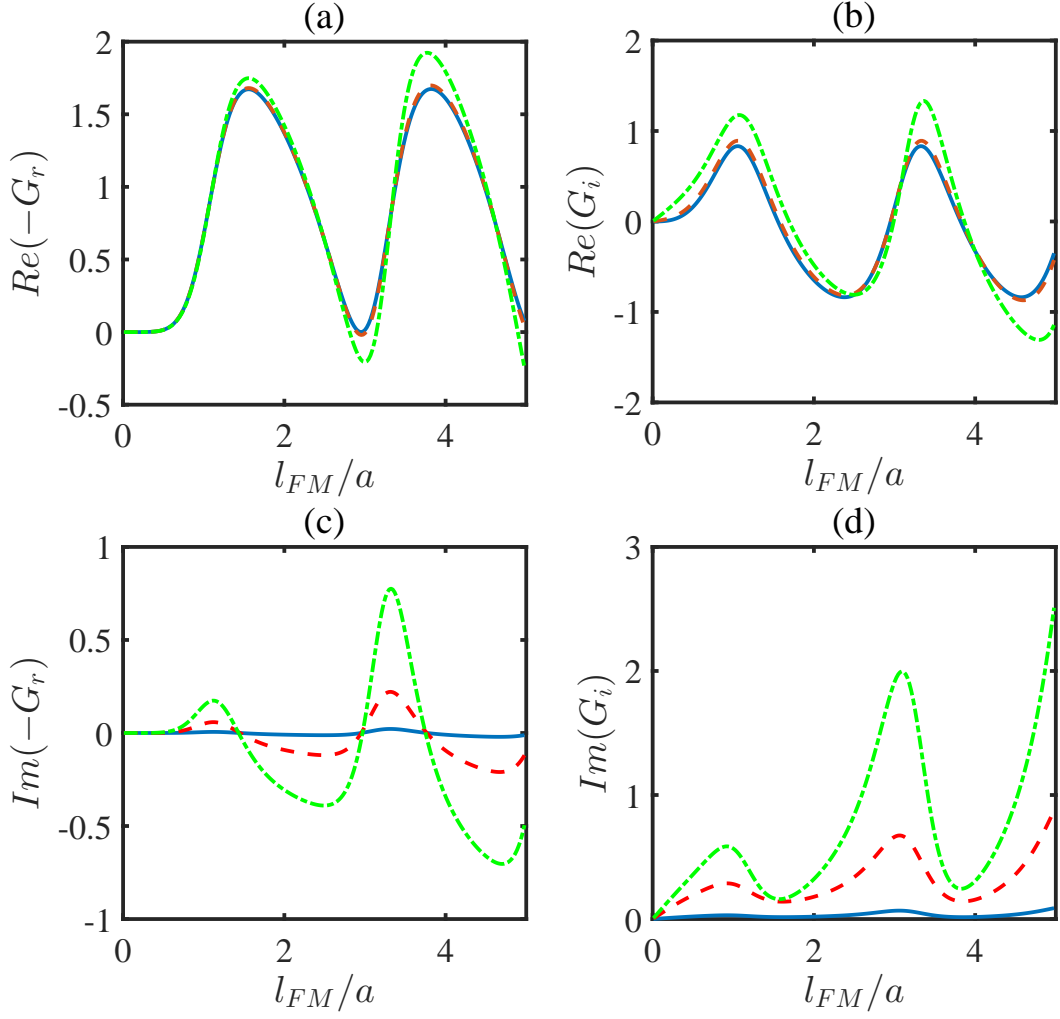}
\caption{\label{FMMGlFM}The numerical results for the real and imaginary parts of $G_{r,i}$ versus the FMM layer thickness $l_{FM}$ with different $\gamma/E$. $\gamma/E=0.01$ for the   solid line, $\gamma/E=0.1$ for the   dashed line, and $\gamma/E=0.3$ for the   dotted line. The other parameters chosen are $\theta=0.3\pi$, $-g S/E=1.2$, and $\mu_0/E=0.01$. The units of  $G_{r,i}$  is $e^2/\hbar a^2$.  }
\end{figure}

 {In Figs.~\ref{FMMGammaS} and \ref{FMMGlFM}, we discuss the influences of the $gS/E$ and $l_{FM}/a$ on the effective spin mixing conductance with different $\gamma/E$ in detail. When the proper system parameters are selected, the spin relaxation $\gamma/E$ will enhance the real parts of effective spin mixing conductance $G_r$ and $G_i$, significantly, which correspond to the traditional definition of spin mixing conductance. This provides us with the possibility to enhance the spin mixing conductance. Meanwhile, the spin relaxation $\gamma/E$ has a strong effect on imaginary parts of $G_r$ and $G_i$, which correspond to the influence of environment on the spin angular momentum.}

\section*{Discussion and Conclusion}
In this work, considering  the ferromagnetic insulator or ferromagnetic metal with spin relaxations described by a non-Hermitian Hamiltonian, we derive an effective spin-transfer torque and an effective spin mixing conductance in the non-Hermitian system. The imaginary parts of the effective spin mixing conductance in the damping-like and field-like direction are no longer zero due to  the spin relaxations. We might divide the effective spin transfer torque into two parts, the change of the macrospin and the spin angular momentum transferred to environment. As an example, we apply the theory to  NM/FMI and NM/FMM bilayer. We found that the spin relaxation has negligible effect on the absolute value of the effective spin conductance of NM/FMI bilayer.
But in NM/FMM, the value of the complex effective spin conductance can be enhanced significantly by the spin relaxations. This provides us with the possibility to enhance the spin
mixing conductance. The dependence of the effective spin mixing conductance on the system parameters (such as insulating gap, \textit{s-d} coupling, and layer thickness) as well as the relations between
the real part and the imaginary part of the effective spin mixing conductance are studied.

\noindent\textbf{Acknowledgments}\\
This work is supported by National Natural Science Foundation of China
(NSFC) under Grants Nos. ~11534002 and 61475033, Subject Construction Project of
School of Physics at Northeast Normal University under Grant No.~111715014, and the China Postdoctoral Science Foundation under Grant No.~2016M600223.

\noindent\textbf{Author Contributions}\\
D. X. Li and  X. X. Yi contributed the idea.
D. X. Li performed the calculations,
and prepared the figures. D. X. Li wrote the main manuscript, H. Z. Shen and H. D. Liu checked the calculations and made an improvement of the manuscript.
All authors contributed to discussion and reviewed the manuscript.

\noindent\textbf{Additional Information}\\
The authors declare no competing financial interests.

\clearpage

\clearpage

\clearpage

\clearpage

\clearpage

\clearpage

\clearpage

\end{document}